\documentclass[12pt,fleqn,notitlepage]{article}
\newtheorem{definition}{Definition}

\newcommand{\blackslug}{\mbox{\hskip 1pt \vrule width 4pt height 8pt 
depth 1.5pt \hskip 1pt}}
\newcommand{\QED}{\quad\blackslug\lower 8.5pt\null\par\noindent}

\newcommand{\cH}{\mbox{${\cal H}$}}

\newcommand{\cR}{\mbox{${\cal R}$}}

\title{Nonstandard numbers for qualitative decision making
\thanks{This work was partially supported 
by the Jean and Helene Alfassa fund for 
research in Artificial Intelligence and by grant 136/94-1 of the 
Israel Science Foundation on ``New Perspectives on Nonmonotonic Reasoning''.}
\\(\small Extended Abstract)
}
\author{Daniel Lehmann\\
Institute of Computer Science, \\Hebrew University, \\Jerusalem 91904, Israel
\\lehmann@cs.huji.ac.il
}
\date{}
\begin{document}
\parindent0.3cm
\maketitle
\begin{abstract}
The consideration of nonstandard models of the real numbers and 
the definition of a
qualitative ordering on those models provides a generalization 
of the principle of
maximization of expected utility. 
It enables the decider to assign 
probabilities of different orders of magnitude to different events
or to assign utilities of different orders of magnitude
to different outcomes.
The properties of this generalized notion of rationality are studied
in the frameworks proposed by von Neumann and Morgenstern 
and later by Anscombe and Aumann.
It is characterized by an original weakening of their postulates 
in two different situations:
nonstandard probabilities and standard utilities on one hand 
and standard probabilities and
nonstandard utilities on the other hand.
This weakening concerns both Independence and Continuity.
It is orthogonal with the weakening proposed by lexicographic orderings.
\end{abstract}
\section{Infinitesimal probabilities}
\label{sec:infpro}
Suppose you are considering playing dice.
You have to choose between betting on six ({\em b6}) 
and betting on four ({\em b4}).
The sums won are the same.
You experiment with the dice and come to the conclusion that the chances
of the dice falling on six are equal to those it falls on four.
You conclude that you are indifferent between {\em b6} and {\em b4}.
You are now offered a third bet ({\em e6}): 
you win if the dice falls on six or falls
on one of the (twelve) edges.
You ponder the chances of the dice falling on an edge and conclude that they
are too small to make you prefer the bet {\em e6} to {\em b4}.
You are indifferent between {\em b4} and {\em e6}. 
Since you believe in the transitivity of
indifference, you conclude you are indifferent between {\em e6} and {\em b6}.

You now consider two more bets.
In {\em e}, you win a large sum if the dice falls on any one of the edges.
In {\em f}, you win the same sum if the dice falls on the edge 
that lies between face six and face five.
You decide you prefer bet {\em e} to bet {\em f}.

According to the theory of rationality proposed 
by von Neumann and Morgenstern~\cite{vonNeuMorg:47},
you are irrational. 
If you are indifferent between {\em e6} and {\em b6}, 
it must be that you give
subjective probability zero to the event of the dice falling 
on one of its edges.
In this case, you {\em must} be indifferent between {\em e} and {\em f}.
You may certainly be rational and indifferent between e and f, 
but must any rational
decider be such?
This work suggests the preferences above 
can be explained by choosing some number
$\epsilon$, positive but infinitesimally close to zero 
(as in Robinson's~\cite{Robinson:66}), 
and assigning a subjective
probability of \mbox{$\frac{1 - \epsilon}{6}$} to the dice falling 
on any of its faces
and a probability of \mbox{$\frac{\epsilon}{12}$} 
to the dice falling on any one of its
edges.
The numbers \mbox{$\frac{1 - \epsilon}{6}$} 
and \mbox{$\frac{1 - \epsilon}{6} + \epsilon$}
are {\em qualitatively} (the term will be formally defined below) equivalent.
The numbers $\epsilon$ and $\frac{\epsilon}{6}$ 
are {\em not} qualitatively equivalent.
\section{Infinitesimal utilities}
\label{sec:infut}
Here is a different example.
Suppose you have to choose between two lotteries.
In the first lottery you may win, with probability $p$,
a week's vacation in Hawaii. 
With probability $1 - p$ you get nothing.
In the second lottery you may win, with the {\em same} probability $p$,
the same vacation in Hawaii, but, with probability $1 - p$ you get
a consolation prize: a free copy of your favorite magazine.
Since the free copy is preferred to nothing, von Neumann-Morgenstern's
independence postulate implies that lottery two is preferred to lottery one.
But couldn't a rational decision maker be indifferent
between the two lotteries?
One, often, I think, buys a lottery ticket in a frame of mind
focused on the big prize and not on the consolation prize.
This behavior is by no means general, as attested by the fact 
lotteries often
offer consolation prizes, but should a decision maker indifferent
between the two lotteries be considered {\em irrational}
in all situations?
I think not.

A variation on this example considers also a third lottery,
in which one wins a trip to Paris with the same probability $p$
as above, and nothing with probability $1 - p$.
Suppose you try to compare lotteries one and three.
You ponder at length the advantages and disadvantages of the
two vacation spots, and decide you are indifferent between the trip to Hawaii
and the trip to Paris, all relevant considerations taken into
account.
You conclude that you are also indifferent between the lotteries one and
three.
The Independence axiom of von Neumann and Morgenstern
implies that lottery two is preferred to lottery three.
But it has been argued that it is quite unreasonable to expect
the very slight improvement that lottery two presents over lottery
one to overcome the lengthy and delicate deliberation that made
you conclude that the trips to Hawaii and Paris are equivalent for you.

Similar examples have been put forward to argue that the indifference 
relation is not always transitive:
lottery three is equivalent to lottery one and to lottery two,
but lottery two is preferred to lottery one.
The system presented in this paper endorses the transitivity
of indifference, but allows a decision maker to be indifferent
between lotteries one and two.

This work proposes to consider that the utility you attach to the trips
to Paris and Hawaii are equal, say $1$. 
The utility you attach to the free copy
of your favorite magazine is $\epsilon$, some positive number, 
infinitesimally close to zero. 
The utility of lottery one and of lottery three is $p$, that of lottery two is
\mbox{$p + (1 - p) \epsilon$}, which is qualitatively equivalent to $p$.
Nevertheless, the free copy of the magazine has utility $\epsilon$, 
that is {\em not}
qualitatively equivalent to zero.
\section{Infinite utilities}
\label{sec:infiniteut}
Let us consider yet another situation.
A patient has to choose between two options:
\begin{enumerate}
\item \label{one} (option $p$) do nothing and die in a matter of weeks,
\item \label{two} (option $q$) undergo surgery, the result of which,
depending on some objective probabilities, may be a long and happy life
(denoted by $l$)
or an immediate death on the operation table (denoted by $d$). 
We shall denote by $\lambda$ the
probability of the surgery being successful, i.e., the probability
of $l$. 
The probability of death on the operation table is, therefore, $1 - \lambda$.
\end{enumerate}
In other terms, one has to compare {\em mixtures} $p$ and
\mbox{$\lambda \, l + (1 - \lambda) \, d$}.
Assuming one prefers $l$ to $p$ and prefers $p$ to $d$,
von Neumann and Morgenstern's postulates imply what we shall
call property {\bf P}: 
there is a unique \mbox{$\lambda \in ]0 , 1[$} for which 
one is indifferent between $p$ and
\mbox{$\lambda \, l + (1 - \lambda) \, d$}.
For any \mbox{$\mu > \lambda$}, one prefers 
\mbox{$\mu \, l + (1 - \mu) \, d$} to $p$, and for all 
\mbox{$\mu < \lambda$}, one prefers $p$ to \mbox{$\mu \, l + (1 - \mu) \, d$}.
If one thinks that a long and happy life is overwhelmingly
preferable to $p$ so as to make the distinction between $p$ and $d$
insignificant, i.e., that
\mbox{$\mu \, l + (1 - \mu) \, d$} is preferable to $p$ 
for any \mbox{$\mu \in ]0 , 1[$},
property {\bf P} fails and 
one deviates from von Neumann-Morgenstern's rationality postulates.

But would we really dismiss as {\em irrational} such a behavior,
or such a preference?
The consideration of an infinite utility for $l$ explains the preference.
Von Neumann-Morgenstern's point of view is perfectly acceptable
and I have no criticism for someone who adheres to it and decides
there is indeed some $\mu$, very close to zero, perhaps, such that
$p$ is equivalent to \mbox{$\mu \, l + (1 - \mu) \, d$}.
My only claim is that someone who thinks 
\mbox{$\mu \, l + (1 - \mu) \, d$} is preferable to $p$ 
for any \mbox{$\mu \in ]0 , 1[$}
cannot be considered {\em irrational} outright.

An argument, very similar to the one just presented,
for prefering a mixture \mbox{$\mu w + (1 - \mu) d$} to
some $p$, for any \mbox{$\mu \in ]0 , 1[$}, even though
$p$ is preferred to $d$, appears in Pascal's~\cite{Pascal:Pensees}.
There $w$ denotes eternal bliss (the reward of the believer
if God exists), $d$ denotes a life spent in error by a believer
in a God that does not exist, and $p$ denotes a life spent
by a non-believer.
This argument, known as Pascal's wager,
is very well-known and the reader may find in a detailed discussion 
in~\cite{Util:Tech2}.

A number of papers~\cite{Brito:75,Shapley:77,Aumann:77} discussed,
in the setting of the St. Petersburg's paradox, the existence of unbounded
utilities.
In the last of these papers, Aumann argues very convincingly
that utilities should be bounded.
At first sight, one may think that infinite utilities imply
unbounded utilities, and therefore Aumann argues also
against infinite utilities, but this is not the case.
His argument may be summarized in the following way:
if utilities were unbounded, for any \mbox{$\lambda \in ]0 , 1[$} 
there would be a consequence ($c$)
such that a lottery 
\mbox{$\lambda \, c + (1 - \lambda) \, d$}
is prefered to a long and happy life ($l$).
But this seems very unreasonable to Aumann.
His argument is directed against {\em unbounded} utilities,
but ineffective against {\em infinite} utilities.
Certainly, no consequence is infinitely preferable to $l$ and therefore,
if there are infinite utilities, the utility of $l$ is infinite.
But there is absolutely no problem if one assumes that
the utility of $l$ is infinite and maybe even maximal 
(nothing is prefered to $l$). In this case,
Aumann's argument disappears. The fact that $l$ is infinitely prefered
to some other consequence, $d$ for example, or a sum of money,
will influence the preferences of a decider between lotteries involving
$l$ and consequences such as $d$.
\section{Background}
\label{sec:back}
Utility theory is discussed in the framework 
of~\cite[Chapter 3]{vonNeuMorg:47}, see also~\cite{Fish:handbook}.
Let $\cH$ be a boolean algebra of subsets of $X$, and $P$ a convex set
of probability measures on $\cH$.We assume that $P$ is finitely generated.
Convex means here that: 
\[
\forall p , q \in P , \: \forall \lambda \in ]0 , 1[, 
\ \lambda p + (1 - \lambda) q \in P.
\]
Here \mbox{$]0 , 1[$} denotes the open real interval in some model, 
maybe nonstandard, 
of the real numbers.

Von Neumann and Morgenstern have characterized the binary relations
$>$ on $P$ that can be defined by a linear functional $u$ on $P$, 
when \mbox{$]0 , 1[$} 
is the standard interval, in 
the following way:
\begin{equation}
\label{eq:fund}
\forall p , q \in P , \ p > q \: \Leftrightarrow \: u(p) > u(q).
\end{equation}
In Equation~\ref{eq:fund}, the functional $u$ is a function from $P$ to
the standard set of real numbers ${\bf R}$ and the relation $>$ in the
right hand side is the usual strict ordering on ${\bf R}$.

Their characterization is equivalent to the following, 
due to Jensen~\cite{Jen:67} 
(see~\cite[p. 1408]{Fish:handbook})
three conditions, for all \mbox{$p , q , r \in P$} and all 
\mbox{$\lambda \in ]0 , 1[$} 
(a weak order is an asymmetric and negatively transitive binary relation):
\[
{\bf A1} \ \ > {\rm \ on \ } P {\rm \ is \ a \ weak \ order} , 
\]
\[
{\bf A2} \ \ \ p > q \: \Rightarrow \: \lambda p + (1 - \lambda) r > 
\lambda q + (1 - \lambda) r,
\]
\[
{\bf A3} \ \ \ (p > q , q > r) \: \Rightarrow \: \exists \alpha , \beta \in ]0 , 1[ ,
{\rm \ such \ that \ } 
\]
\[
\ \ \ \ \ \ \ \ \ \alpha p + (1 - \alpha) r > q > \beta p + (1 - \beta) r.
\]
The three conditions above are not the original postulates of
von Neumann and Morgenstern, they are equivalent to them.
They will be referred to, nevertheless, in this work, as 
von Neumann and Morgenstern's postulates.
The purpose of this work is to generalize von Neumann and Morgenstern's
characterization to deal with {\em qualitative} probabilities or with 
{\em qualitative} utilities.
In the sequel, \mbox{$p \geq q$} will denote \mbox{$q \not > p$} and
\mbox{$p \sim q$} will denote the conjunction of \mbox{$p \geq q$} and
\mbox{$q \geq p$}.
\section{Qualitative Decision Theory}
Qualitative decision theory has been developed mostly in opposition to
quantitative decision theory, stressing decision methods that do not 
satisfy von Neumann-Morgenstern's or Savage's~\cite{Savage:54} postulates, 
the postulates generally accepted for quantitative decision theory.
The focus in qualitative decision theory has always been on methods
and algorithms, more than on an axiomatic treatment (\cite{BrafTen:qdt}
is an exception).

A different approach is proposed here: qualitative and quantitative
decision theories can be viewed as special cases of a unified general
theory of decision that contains both.
This unified theory is a generalization of the quantitative theory.
The power of the generalization lies in the consideration of nonstandard
models of the set of real numbers for utilities
and a definition of indifference that neglects infinitesimally small
differences.
In this paper probabilities will always be standard.
Some first results, for nonstandard probabilities and standard utilities
have been presented in~\cite{Leh:UAI96}.
Preliminary ideas appeared in~\cite{Leh:QDT}.

A well-established tradition in Decision Theory considers Expected Utility
Maximization as the only rational policy.
Following this view, an act $f$ is strictly preferred to an act $g$
iff the utility expected from $f$ is strictly larger than that
expected for $g$.
Since expected utilities are real numbers, {\em strictly larger} has
its usual, {\em quantitative} meaning.
The main claim of this paper is that the qualitative point of view
may be subsumed by a slightly different definition of {\em strictly larger}.
Suppose we consider any model elementarily equivalent to the real numbers,
more precisely, any (standard or nonstandard) model of the real numbers,
$\cR$ (for the standard model, we shall use ${\bf R}$).
Let $x$ and $y$ be elements of $\cR$. 
To make matters simpler, suppose that both $x$ and $y$ are positive.
The number $x$ is quantitatively
larger than $y$ iff \mbox{$x - y > 0$}.
What could be a reasonable definition of {\em qualitatively larger}?
Clearly, if $x$ is qualitatively larger than $y$ then it must be quantitatively
larger: in a sense {\em qualitatively larger} means {\em definitely larger}.
A first idea that may be considered is to use a notion that proved
fundamental for nonstandard analysis (the monads of~\cite{Robinson:66},
or see~\cite{Keisler:76}):
the notion of two numbers being {\em infinitely close}, 
and consider that a number
$x$ is {\em qualitatively}
larger than a number $y$ iff $x$ is larger than $y$ and {\em not
infinitely close} to $y$, i.e., iff \mbox{$x - y$} is 
strictly larger than some positive standard number.
At first this idea looks appealing:
if $\epsilon$ is strictly positive and infinitesimally close to zero,
and $x$ is a standard, strictly positive, real,
we do {\em not} want \mbox{$x + \epsilon$} to be qualitatively larger than
$x$.
At a second look, one realizes that the size of \mbox{$x - y$} should not
be judged absolutely, but relatively to the size of $x$: for example
\mbox{$\epsilon^{2} + \epsilon$} should be qualitatively larger than
$\epsilon^{2}$. Therefore I propose the following definition:
\begin{definition}
\label{def:qual>}
Let $x$ and $y$ be positive. We shall say that $x$ is 
qualitatively larger than $y$ and write \mbox{$x \succ y$} iff 
\mbox{${x - y} \over {x}$} is strictly positive and {\em not} infinitesimally
close to zero:
in other terms, iff there is a strictly positive {\em standard} number
$r$ such that \mbox{${{x - y} \over {x}} \geq r$}.
\end{definition}
The definition may be extended to arbitrary numbers in an obvious way:
\begin{enumerate}
\item
if \mbox{$x \geq 0$} and \mbox{$0 > y$}, then \mbox{$x \succ y$}, and
\item
\mbox{$x \succ y$} iff \mbox{$-y \succ -x$},
\end{enumerate} 

\mbox{$x \preceq y$} shall denote \mbox{$ x \not \succ y$} 
and \mbox{$x \sim y$} shall denote that 
\mbox{$x \preceq y$} and \mbox{$y \preceq x$}.

Notice that, if we choose, for $\cR$, the standard model of the reals, 
${\bf R}$,
then \mbox{$x \succ y$} iff \mbox{$x > y$}.
Therefore our treatment would include the classical approach, 
if we allowed also negative utilities. As said above, in this paper,
we concentrate on the case all utilities are positive.
Is our framework, with positive nonstandard utilities, a generalization
of the classical theory, with standard positive and negative utilities?
Since, in the classical setting, utilities are defined only up
to an additive constant, bounded utilities may always be
considered to be positive, 
by adding a positive large enough constant.
In view of Aumann's~\cite{Aumann:77} critique of unbounded utilities,
we feel that the present framework encompasses the most important
part of classical theory.

Notice also that Definition~\ref{def:qual>} relies on the notion
of a {\em nonstandard} number, and that notion is not first-order definable.

Expected Qualitative Utility Maximization, the paradigm of rationality proposed
here is the version of
Expected Utility Maximization that obtains when, 
for probabilities and utilities, 
\begin{itemize}
\item the models chosen for the real numbers {\em may} be nonstandard, and
\item real numbers are compared {\em qualitatively}, i.e., by $\succ$.
\end{itemize}
At this stage, I do not know of an axiomatic characterization
of Expected Qualitative Utility Maximization in its most general form:
nonstandard probabilities and utilities.
But two orthogonal special cases have been characterized in full:
first, the case in which probabilities may be nonstandard but utilities
are standard and secondly, the case in which probabilities are standard 
but utilities may be
nonstandard.
In the first case, we want to characterize the binary relations
$>$ on $P$ that can be defined by a functional 
\mbox{$u: P \rightarrow {\bf R}$} into the standard real numbers in 
the following way:
\begin{equation}
\label{eq:st}
\ \ \ \ \forall p , q \in P , p > q \Leftrightarrow u(p) > u(q), 
\end{equation}
and $u$ is pseudo-linear, i.e.:
\begin{equation}
\label{eq:pseudo}
\ \ \ \ \forall p , q \in P , \forall \lambda \in ]0 , 1[ , 
u(\lambda p + (1 - \lambda) q) \sim \lambda u(p) + (1 - \lambda) u(q).
\end{equation}
In Equation~\ref{eq:pseudo}, the interval \mbox{$]0 , 1[$} may be non-standard
and therefore the right-hand side of the equivalence may be nonstandard.
This case is treated in Section~\ref{sec:post1}.

In the second case, we characterize the binary relations
$>$ on $P$ that can be defined by a {\em linear} functional 
\mbox{$u: P \rightarrow \cR_{+}$} in 
the following way:
\begin{equation}
\label{eq:nonst}
\ \ \ \ \forall p , q \in P , p > q \Leftrightarrow u(p) \succ u(q).
\end{equation}
Here the interval \mbox{$]0 , 1[$} is the standard one.
This case in treated in Section~\ref{sec:post2}.

One should immediately notice that, 
if \mbox{$c > 0$}, the utility function $c \: u(p)$ defines the same
ordering as $u(p)$, and is linear or pseudo-linear iff $u$ is.
But, contrary to what happens in the classical setting, if $d \in \cR$
the function \mbox{$d + u(x)$} does not, in general, define the same
ordering as $u(p)$.
Such an instability under an additive constant, and in particular
an asymmetry between gains and losses has
been found in the behavior of decision makers 
in many instances~\cite{FishKoch:79,KahnTver:79,Schoe:80}.
The question of whether Expected Qualitative Utility Maximization
is a realistic model for explaining such behavior cannot be discussed
in this work.
\section{Maximin as Expected Qualitative Utility Maximization}
Considering nonstandard utilities enables us
to obtain decision criteria that do not satisfy 
von Neumann-Morgenstern's postulates and were so far considered as
part of the realm of {\em qualitative} decision theory.

As noticed above, Expected Qualitative Utility Maximization generalizes
Expected Utility Maximization: if one chooses the standard model for
real numbers then Expected Qualitative Utility Maximization 
boils down exactly to Expected Utility Maximization, at least when
utilities are bounded.
We shall show now that considering nonstandard utilities enables us
to obtain decision criteria that do not satisfy 
von Neumann-Morgenstern's postulates and were so far considered as
part of the realm of {\em qualitative} decision theory.
A version of the Maximin criterion will be presented.
The Maximin criterion has been proposed by A. Wald~\cite{Wald:50},
in a different framework. The criterion to be presented is
a variation on this theme.

Assume the set $X$ is finite and $\cH$ contains all subsets of $X$.
Let the elements of $X$ be \mbox{$x_{0}, \ldots, x_{n - 1}$}.
Let $\epsilon$ be a number that is positive and infinitesimally close to zero
and let our utility function $u$ be the linear function defined by:
\mbox{$u(x_{i}) = \epsilon ^ {n - i - 1}$}, 
for \mbox{$i = 0 , \ldots , n - 1$}.
Notice that \mbox{$x_{i} < x_{j}$} iff \mbox{$i < j$}.
The utility of a mixture 
\mbox{$\lambda x_{i} + (1 - \lambda) x_{j}$} is
\mbox{$\lambda \epsilon ^ {n - i - 1}$} if \mbox{$x_{i} < x_{j}$} and
\mbox{$\epsilon^{n - i - 1}$} if \mbox{$x_{i} \sim x_{j}$}.

Suppose \mbox{$i < j$} and \mbox{$i ' < j '$}.
Then, 
\mbox{$\lambda x_{i} + (1 - \lambda) x_{j} < \mu x_{i '} + 
(1 - \mu) x_{j '}$} iff \mbox{$x_{i} < x_{i '}$}
or \mbox{$x_{i} \sim x_{i '}$} and \mbox{$\lambda > \mu$}.
The decision maker therefore compares different mixtures by comparing
the worst possible outcomes and, if they are the same, their respective
probabilities. This is some form of Maximin criterion
and does not satisfy {\bf A2} or {\bf A3}, but it has been considered
a rational way of deciding by many authors, and it
is amenable to Expected Qualitative Utility Maximization.
\section{Postulates for nonstandard probabilities and standard utilities}
\label{sec:post1}
The postulates that characterize this first case are {\bf A1}, {\bf A3} and the following {\bf B2}.
\begin{definition}
\label{def:al}
\mbox{$\lambda \in ]0 , 1[$} is {\em negligible} iff, for any \mbox{$p , q \in P$},
\mbox{$\lambda \: p + (1 - \lambda) \: q \sim q$}.
\end{definition}
The intuitive meaning of {\em negligible} is infinitesimally close to zero.
\[
{\bf B2} \ \ p > q , \lambda {\rm \ not \ negligible \ } \Rightarrow  
\ \lambda p + (1 - \lambda) r > \lambda q + (1 - \lambda) r.
\]
\section{Postulates for standard probabilities and nonstandard utilities}
\label{sec:post2}
The postulates that characterize this second case are {\bf A1} and the following.

To formulate our independence property, it is best to set 
the following definition.
\begin{definition}
\label{def:>>}
We shall say that $p$ overrides $q$ and write \mbox{$p \gg q$} iff
\mbox{$p > q$} and for any $q'$ such that \mbox{$q > q'$}
and for any \mbox{$\lambda \in ]0 , 1[$},
\mbox{$\lambda q + (1 - \lambda) p \sim \lambda q' + (1 - \lambda) p$}.
\end{definition}
The intuitive meaning of \mbox{$p \gg q$} is that $p$ is so much preferred
to $q$ that, in any lottery in which $p$ and $q$ are the
prizes, if one does not win $p$, one does not even care to cash $q$, 
but would as well get any lesser prize $q'$.
Notice that, since $>$ is asymmetric, the relation $\gg$ is also
asymmetric and therefore irreflexive.

Our independence property may now be formulated as:
\[
{\bf A'2} \ \ p > q , r \not \gg p \: \Rightarrow \: 
\forall \lambda \in ]0,1[  
\ \lambda p + (1 - \lambda) r > \lambda q + (1 - \lambda) r.
\]
The intuitive meaning of {\bf A'2} is that any lottery is sensitive to
both its prizes, unless one of the prizes overrides the other one.

\[
{\bf A'3} \ \ p > q > r \: \Rightarrow \: 
\exists \alpha \in ]0 , 1[ {\rm \ such \ that \ }
\alpha p + (1 - \alpha) r > q .
\]
\[
{\bf A''3} \ \ p > q > r , p \not \gg q \: \Rightarrow \: 
\exists \beta \in ]0 , 1[ {\rm \ such \ that \ }
q > \beta p + (1 - \beta) r.
\]
\section{Comparison with previous work}
Numerous works during the fifties and the sixties considered
weakenings of the von Neumann-Morgenstern's postulates.
Nonstandard analysis~\cite{Robinson:66} appeared late on the scene.
This work proposes an original weakening based on
nonstandard analysis.

Our postulates are very close to the original postulates
of von Neumann and Morgenstern. In particular
the ordering $<$ is modular (weak total) 
and the indifference relation
$\sim$ is transitive. 
In the case of nonstandard utilities, we weaken both {\bf A2} and {\bf A3},
in a closely linked manner.
Notice that {\bf A2}, {\bf A'3} and {\bf A''3} together
imply {\bf A3}, since {\bf A2} says that \mbox{$p \gg q$}
implies that for any $w$, \mbox{$w \geq q$}.
The lexicographic orderings of~\cite{BlumeBranDek:91} provide
one of the best known weakenings of von Neumann and Morgenstern's postulates.
The weakening they present is essentially orthogonal to ours.
Indeed, the lexicographic orderings
define a preference relation that satisfies {\bf A2}.
Any ordering that satisfies our postulates and those of
von Neumann-Morgenstern postulates.
To explain better the difference between lexicographic orderings
and our qualitative ordering (in the case probabilities are standard), 
assume $P$ is the real plane ${\bf R}^{2}$
ordered by the lexicographic ordering: \mbox{$(x , y) < (x' , y')$}
iff either \mbox{$x < x'$} or \mbox{$x = x'$} and \mbox{$y < y'$}.
For \mbox{$\lambda \in [0 , 1]$}, define 
\mbox{$\lambda (x , y) + (1 - \lambda) (x' , y')$}
to be \mbox{$(\lambda x + (1 - \lambda) x' , \lambda y + (1 - \lambda) y')$}.
Notice that \mbox{$(0 , 0) < (1 , 10) < (2 , 0)$}, but there is no 
\mbox{$\lambda \in ]0 , 1[$} such that 
\mbox{$(1 , 10) =$} \mbox{$\lambda (0 , 0) + (1 - \lambda) (2 , 0)$}
since \mbox{$\lambda (0 , 0) + (1 - \lambda) (2 , 0) =$}
\mbox{$(2 (1 - \lambda) , 0)$}.
Both lexicographic and qualitative orderings imply the failure
of property {\bf P} of section~\ref{sec:infiniteut}.
But lexicographic orderings also implies the failure of the following
property that holds in the case probabilities are standard:
if \mbox{$p > q > r$} and
there exists some \mbox{$\beta \in ]0 , 1[$}
such that
\mbox{$q > \beta p + (1 - \beta) r$}, then there
exists some \mbox{$\gamma \in ]0 , 1[$}
such that
\mbox{$q \sim \gamma p + (1 - \gamma) r$}. 
Indeed there exists some \mbox{$\beta \in ]0 , 1[$}
such that
\mbox{$(1 , 10) >$}\mbox{$\beta (2 , 0) + (1 - \beta) (0 , 0) =$}
\mbox{$\beta (2 , 0)$}:
for example \mbox{$\beta = 0.4$}, and nevertheless there is no $\lambda$
as above.
Lexicographic and qualitative orderings stem from different concerns
and have very different characteristics.
\section{Subjective probability}
\label{sec:Subj}
\subsection{Anscombe-Aumann's framework}
\label{sub:AnscAum}
In~\cite{AnscAum:63}, Anscombe and Aumann consider
a finite set $S$ (of states) and the set ${\bf F}$ (of acts)
of mappings: \mbox{$S \mapsto P$}.
They show that a single postulate , added to {\bf A1}, {\bf A2} 
and {\bf A3} is enough to characterize the orderings obtainable 
from subjective probabilities on states and linear utilities:
for any \mbox{$a , b \in {\bf F}$},
\[
{\bf A4} \ {\rm \ If \ } \forall s \in S , s \neq s_{0} 
\Rightarrow a(s) = b(s), {\rm \ then \ }
a > b \: \Rightarrow \: a(s_{0}) > b(s_{0}).
\]
In the last part of {\bf A4}, \mbox{$a(s_{0})$} and
\mbox{$b(s_{0})$} stand for the corresponding constant functions.

For the case of standard utilities (and nonstandard probabilities),
the same single added postulate is enough to guarantee 
the corresponding result: existence of nonstandard subjective probabilities and
a pseudo-linear utility function into the standard reals.

For the case of standard probabilities (and nonstandard utilities), in addition
to {\bf A1}, {\bf A'2}, {\bf A'3}, {\bf A''3} and {\bf A4}, one needs
an additional postulate to ensure the subjective probabilities are standard.
This postulate deals with Savage-null states.
\[
{\bf A'5} \ \ \ t \in S , \: a \in {\bf F} , \: a(t) \gg a \: \Rightarrow \:
t {\rm \ is \ null}.
\]
where null is defined below and $\gg$ is defined in Definition~\ref{def:>>}.
\begin{definition}
\label{def:Lzero}
Let \mbox{$t \in S$}. The element $t$ is said to be {\em null}
iff for any \mbox{$a , b \in {\bf F}$} we have \mbox{$a \sim b$}
if $a$ and $b$ agree everywhere except possibly on $t$, i.e.,
for any \mbox{$s \neq t$}, \mbox{$a(s) = b(s)$}.
\end{definition}
\section{Conclusion}
Nonstandard models of the real numbers provide for a natural notion of
qualitative equivalence and a principle of Qualitative Utility Maximization.
This work characterizes in full the situation in which one allows 
nonstandard utilities but insists on standard probabilities.
In this framework one may consider consequences that are infinitely
preferable to others and criteria of the Maximin family.
The study of games with nonstandard utilities seems appealing.
The dual case of nonstandard probabilities and standard utilities
and the most general case of nonstandard probabilities and utilities
are yet to be characterized.
They will include consideration of subjective probabilities
infinitesimally close to zero.
\section*{Acknowledgments}
I want to thank Israel Aumann for his encouragements
and suggestions to improve the presentation of this work.
\bibliographystyle{plain}

\end{document}